\documentclass{aa}
\usepackage{graphicx,epsfig,rotating}
\def\Agata{R\' o\. za\' nska~}

\begin{document}

               
\title{The structure and X-ray radiation spectra of illuminated accretion 
 disks in AGN.\ III. Modeling fractional variability}

\author{R.~W.~Goosmann \inst{1,2} \thanks{Email: goosmann@astro.cas.cz},
        B.~Czerny \inst{2,3} \thanks{Email: bcz@camk.edu.pl},
        M.~Mouchet \inst{2,4} \thanks{Email: martine.mouchet@obspm.fr},
        G.~Ponti \inst{5,6,7} \thanks{Email: ponti@bo.iasf.cnr.it},\\
        M.~Dov\v{c}iak \inst{1} \thanks{Email: dovciak@astro.cas.cz},
        V.~Karas \inst{1} \thanks{Email: vladimir.karas@cuni.cz},
        A.~R\'o\.za\'nska \inst{2,3} \thanks{Email: agata@camk.edu.pl}, \and 
        A.-M.~Dumont \inst{2} \thanks{Email: anne-marie.dumont@obspm.fr}
}
\offprints{R.~W.~Goosmann (goosmann@astro.cas.cz)}
\institute{
$^1$~Astronomical Institute, Academy of Sciences of the Czech Republic,
     Bo{\v c}ni II 1401, 14131 Prague, Czech Republic\\
$^2$~Observatoire de  Paris, Section de Meudon, LUTH, 5 place Jules Janssen,
     92195 Meudon Cedex, France\\
$^3$~Copernicus Astronomical Center, Bartycka 18, 00-716 Warsaw, Poland\\
$^4$~Laboratoire Astroparticules et Cosmologie, Universit\'e Denis
     Diderot, 2 place Jussieu, 75251 Paris Cedex 05, France\\
$^5$~Dipartimento di Astronomia, Universit\`a di Bologna, Via Ranzani 1,
     I--40127, Bologna, Italy\\
$^6$~INAF--IASF Sezione di Bologna, Via Gobetti 101, I--40129, Bologna, Italy\\
$^7$~Institute of Astronomy, Madingley Road, Cambridge CB3 0HA, United Kingdom}

\authorrunning{Goosmann et al.}

\titlerunning{Structure and X-ray spectra of AGN accretion disks III:
  rms-variability}

\date{Received ...; accepted ...}

\abstract{Random magnetic flares above the accretion disks of Active Galactic
Nuclei can account for the production of the primary radiation and for the
rapid X-ray variability that have been frequently observed in these
objects. The primary component is partly reprocessed in the disk atmosphere,
forming a hot spot underneath the flare source and giving rise to distinct
spectral features. Extending the work of Czerny et al. (2004), we model the
fractional variability amplitude due to distributions of hot spots co-orbiting
on the accretion disk around a supermassive black hole. We compare our results
to the observed fractional variability spectrum of the Seyfert galaxy
MCG-6-30-15. According to defined radial distributions, our code samples
random positions for the hot spots across the disk. The local spot emission is
computed as reprocessed radiation coming from a compact primary source above
the disk. The structure of the hot spot and the anisotropy of the re-emission
are taken into account. We compute the fractional variability spectra expected
from such spot ensembles and investigate dependencies on the parameters
describing the radial spot distribution. We consider the fractional
variability $F_{\rm{} var}$ with respect to the spectral mean and the
so-called point-to-point definition $F_{\rm{} pp}$. Our method includes
relativistic corrections due to the curved space-time in the vicinity of a
rotating supermassive black hole at the disk center; the black hole's angular
momentum is a free parameter and is subject to the fitting procedure. We
confirm that the rms-variability spectra involve intrinsic randomness at a
significant level when the number of flares appearing during the total
observation time is too small. Furthermore, the fractional variability
expressed by $F_{\rm{} var}$ is not always compatible with $F_{\rm{} pp}$. In
the special case of MCG-6-30-15, we can reproduce the short-timescale
variability and model the suppressed variability in the energy range of the
K${\alpha}$ line without any need to postulate reprocessing farther away from
the center. The presence of the dip in the variability spectrum requires an
increasing rate of energy production by the flares toward the center of the
disk. The depth of the feature is well represented only if we assume a fast
rotation of the central black hole and allow for considerable suppression of
the primary flare emission. The modeled line remains consistent with the
measured equivalent width of the iron K$\alpha$ line complex. The model can
reproduce the frequently observed suppression of the variability in the
spectral range around 6.5 keV, thereby setting constraints on the black hole
spin and on the disk inclination.

}

\maketitle

\section{Introduction}

The variability of the X-ray emission is one of the basic characteristics  of
Active Galactic Nuclei (AGN). Well established since the EXOSAT observations
(Pounds 1985), the variability is expected to be an important tool in the
study of the accretion process, although its aperiodic noise-like character was
early on recognized as a source of interpretation difficulty (McHardy \&
Czerny 1987; Lawrence at al. 1987).

The phenomenological approach to X-ray variations has brought significant
information on the properties of the lightcurves, such as their power spectral
density (e.g. Markowitz et al. 2003;  McHardy et al. 2005) and their process
nonlinearity (Uttley et al. 2005). Nevertheless, the physical processes behind
the observed variability have not yet been fully understood. The same is true
for accreting galactic black holes. The main source of the ambiguity is that
several different models can account for the average radiation spectra,
e.g. the 'hot inner flow model' (Ichimaru 1977; Narayan \& Yi 1994; Kato et
al. 2004; Sobolewska et al. 2004a), the 'continuous corona model' (Liang \&
Price 1977; Bisnovatyi-Kogan \& Blinnikov 1977; Czerny \& Elvis 1987; Haardt
\& Maraschi 1991; Czerny et al. 2003; Merloni 2003), the 'patchy corona model'
(Galeev et al. 1979; Haardt et al. 1994; Stern et al. 1995; Ross et al. 1999;
Nayakshin et al. 2000; Sobolewska et al. 2004b), or the 'jet-basis model'
(Henri \& Pelletier 1991; Markoff et al. 2005). Several models can also
account for the power density spectra, e.g. the 'shot model' (Lehto 1989),
the 'self-organized critical state model' (Mineshige et al. 1994), or the
'traveling perturbations model' (Lyubarskii 1997; King et al. 2004).
Apparently,  it is impossible to uniquely determine the accretion flow
geometry on the basis of power spectra or from the analysis of the mean
spectral shape alone.
 
Therefore, a 2-D approach is needed, which would combine the time and the
energy dependence. Such an  approach was pioneered in the context of binary
black holes using various techniques, e.g., phase lags (Miyamoto et al. 1988;
Cui et al. 1997), correlation functions (Nolan et al. 1981; Poutanen \& Fabian
1999), coherence functions (Vaughan \& Nowak 1997), frequency resolved spectra
(Gilfanov et al. 2000; Zycki 2002; Zycki 2003), wavelet analysis (Lachowicz \&
Czerny 2005). Now the appropriate data is starting to be available for AGN
(e.g. Yaqoob et al. 2003; Kaastra et al. 2004; Ponti et al. 2004; Markowitz
2005). We therefore concentrate on modeling the energy-dependent fractional
variability amplitude (Vaughan et al. 2003). Long light curves obtained from
Chandra and XMM-Newton satellites allow us to determine this function for a
few AGN. Such detailed information provides a challenge to the models.

In the present paper, we consider the magnetic flare model of the X-ray
emission, suggested by Galeev et al. (1979) in the context of Cyg X-1, and
further developed in various directions by several authors, both in the context
of galactic sources (Poutanen \& Fabian 1999; Beloborodov 1999)  and AGN
(Haardt et al. 1994; di Matteo 1998; Merloni \& Fabian 2001;
Torricelli-Ciamponi et al. 2005). In this picture, the magneto-rotational
instability (MRI) operating in the disk occasionally leads to the rise of
large magnetic loops high above the disk surface. Field reconnection occurs in
this region and leads to the creation of localized regions of hot plasma
(flares). The flares are cooled by Comptonizing the disk emission, which
produces a power law X-ray radiation, customarily called primary emission. A
fluctuating number of flares depending on various parameters is supposed to be
responsible for the observed AGN variability. Recent 3-D simulations of the
magneto-rotational  instability within the disk seem to support such a picture
(e.g. Miller  \& Stone 2000; Turner 2004), but simulations including the
Compton cooling are still to be done. 

The hard X-ray emission of a flare leads to the creation of  a hot spot at the
disk surface where the X-ray emission is  reflected/reprocessed. The combined
emission of all spots thus constitutes the  so-called reflected component in
the total X-ray spectrum, and the issue of the reprocessing of the incident
X-ray radiation by the disk surface has been extensively studied
(e.g. Lightman \& White 1988;  Ross \& Fabian 1993; Zycki et al. 1994;
Nayakshin et al. 2000; Ballantyne et al. 2001; \Agata et al. 2002).

Here, we extend our previous work on this subject (Czerny et al. 2004). We
improve our description of the spot emission by taking its dependence on the
distance from the spot center and on the local emission direction into
account. We use these new computations to model the standard fractional
variability amplitude as well as the point-to-point fractional variability
amplitude. We compare the results to observational data of MCG-6-30-15.

\section{The model}
\label{sect:model}

\subsection{Radiative transfer within the spot}
\label{sect:transfer}

Let the flare be located at a certain height above an accretion disk. It
illuminates the disk surface. We assume that the flare source is very compact
and approximate it by an isotropic, point-like source. The emitted spectrum is
of a power law shape, with the photon index $\Gamma = 1.9$, extending from 1
eV to 100 keV.

\begin{figure}
  \epsfxsize=8.8cm
  \epsfbox{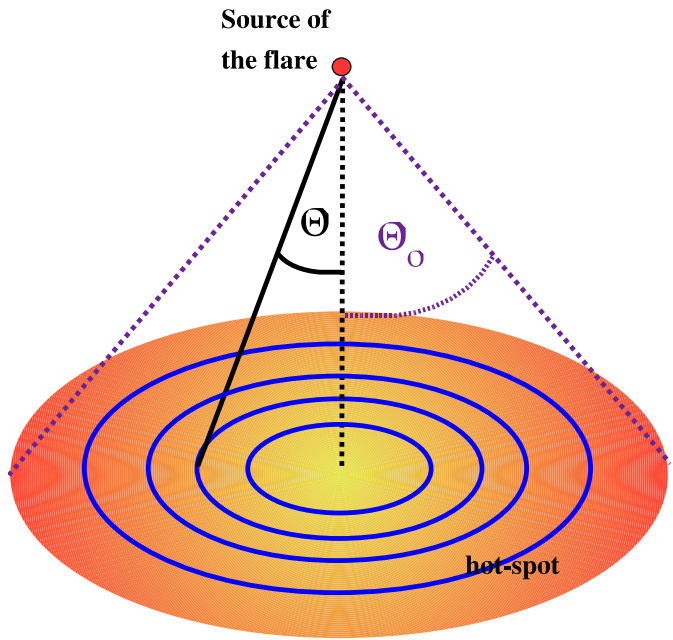}
  \caption{Spot approximated by five rings below a flare. In each of the
  rings, the reflected spectrum is calculated for the corresponding incident
  angle $\theta$ of the X-ray radiation and the ionization parameter at the
  surface.}
  \label{fig:rings}
\end{figure}

The reprocessed emission depends on the distance from the flare. Therefore, we
divide the patch into concentric rings (see Fig.~\ref{fig:rings}). Within each
ring, the incident angle of the hard X-ray illumination and irradiating flux
is approximately constant. It is convenient not to parameterize the rings by
their radial distance from the projected position of the flare (spot-center),
but instead by  $\mu$, which is the cosine of the incident angle $\theta$ of
the irradiating flux (see Fig.~\ref{fig:rings}). We vary $\mu = \cos{\theta}$
in constant steps of 0.1, ensuring that all rings intercept the same fraction
of the total illuminating flux.

The reflected spectrum of each ring is calculated for five values of the local
emission angle, $\psi$, measured with respect to the disk axis. We choose
equal steps in $\cos{\psi}$ from 0.1 to 0.9. The local radiation spectrum is
computed accurately by performing radiative transfer simulations in
plane-parallel approximation. The vertical structure of the illuminated disk
is computed using the code of \Agata et al. (2002). Following an argument made
by Collin et al. (2003), we assume that the total flare duration is
significantly shorter than the characteristic timescale for the restoration of
the hydrostatic equilibrium in the suddenly irradiated disk. We thus use the
vertical density profile of the disk as computed before the flare turns on.

The local radiative transfer computations are conducted by coupling the codes
{\sc titan} and {\sc noar}, as described in Dumont et al. (2000) and updated in
Dumont et al. (2003). To model the accretion disk atmosphere, we assume a
vertically stratified, plane-parallel medium being irradiated by hard X-ray
radiation from above its upper surface. From underneath, low-temperature
black-body radiation is added to account for the heating of the underlying
disk. The computations are performed for a $10^8 M_{\odot}$ black hole with a
disk accretion rate of 0.001 of the Eddington accretion rate. A distance of 18
$R_{\rm{} g}$ ($R_g = GM/c^2$) from the black hole is assumed, and the ratio of
the incident flux to the disk flux equals 144 at the center of the
spot. Details of these calculations are described in Goosmann et al. (2006).

The overall response of the irradiated material to the incident flux is in
agreement with known results: a hot layer forms at the top of the disk
atmosphere, that is roughly at the inverse Compton temperature, followed by a
steep transition to colder, less ionized layers. Such behavior is due to the
requirement of the hydrostatic equilibrium (Field 1965; Krolik et al. 1981;
Raymond 1993; Ko \& Kallman 1994; \Agata \& Czerny 1996; Nayakshin et
al. 2000; Ballantyne et al. 2001; \Agata et al. 2002). However, the horizontal
stratification of the spot introduced in our model brings new aspects to the
resulting reprocessing: we include the fact that the disk surface is most
strongly irradiated directly below the flare, while the irradiating flux
decreases and the incident angle of the irradiation increases with the
distance from the hot spot center. Therefore, the surface inside a hot spot
shows a significant radial gradient of its properties, including the locally
emitted spectrum. The incident angle and the local incident flux are different
in each ring. Large differences between the ring spectra are seen in the soft
X-ray band. Spectra from the outer rings are much harder than those from the
inner rings, since the reflector there is less ionized. The high-energy parts
of the local spectra from all rings are rather similar to each other.

\subsection{Ring integration and dependence on the local emission angle}
\label{sect:integ}

\begin{figure}
  \epsfxsize=8.8cm
  \epsfbox{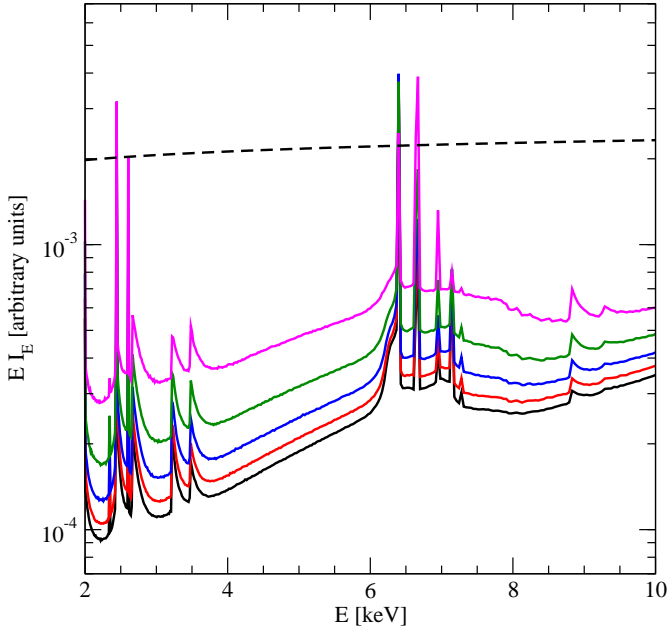}
  \caption{The spot spectra (intensity) integrated over the rings for 5 values
    of the emission angle $\psi$: $\cos \psi =$ 0.1, 0.3, 0.5, 0.7, and
    0.9. The upper spectrum is the side view. Parameters adopted:
    $M=10^8M_{\odot}$, $R = 18 \; R_{\rm{} g}$, $\dot m_{disk}= 0.001$, and
    $L_{\rm{} X}/L_{\rm{} d} = 144$.}
  \label{fig:whole_spot}
\end{figure}

The final spectrum of a spot, for a given local emission direction, is
determined by the integration over the rings. Due to the adopted
parameterization of the rings, the integration actually reduces to a simple
summation. If the flare emission is somehow collimated toward the disk, the
outer parts of the hot spot are not irradiated. Therefore, we can treat the
opening angle of the irradiating cone as an additional parameter of the
model. Here, we adopt a half-opening angle $\Theta_0 = 60^{\circ}$. This value
of $\Theta_0$ corresponds to half of the illuminating flux being reprocessed
by the disk medium and also ensures that our calculations remain accurate with
respect to the local conditions of the assumed hydrostatic equilibrium. The
resulting local spectra are shown in  Fig.~\ref{fig:whole_spot}. These
$\psi$-dependent spot spectra are further adopted as representative of
computations of a whole accretion disk covered by many spots.

We note that by integrating over the spot surface, we average the
spectrum and lose some spatial resolution. We do that in the present
model for simplicity. In reality, the properties of the incident and
reprocessed radiation change across the spot, which affects the single
spot lightcurve. An investigation of this effect is part of a detailed
study about individual flare spots (Goosmann et al. 2006). From this
study, it follows that in our case the integration over the spot
surface does not induce a major error. The intensity of the local
spectra at a given emission angle $\psi$ changes regularly across the
spot surface and only within a factor of $\sim 2-3$. Also, the
spectral slope only changes by a factor $<2$.

The shape of the radiation spectrum of a whole spot does not depend strongly
on the emission angle $\psi$, while the overall normalization does. The
normalization rises toward higher values of $\psi$, which represents a
limb-brightening effect. The spectrum for all emission angles are much harder
than the local spot spectrum used by Czerny et al. (2004). This is caused by
our present assumption of short lasting flares, while Czerny et al. (2004)
assumed that the flare duration is long enough for the underlying irradiated
disk layers to settle into a new hydrostatic equilibrium.

\subsection{Flare distribution across the disk}
\label{sec:flare-dist}

To define the flare distribution across the disk, we generally follow
the simple approach outlined in Czerny et al. (2004). We assume that the
spectral shape of all flares is the same, since it consumes too much computer
time to repeat such complex computations separately for spots located
at various disk radii. On the other hand, we take the fact that the
reflected (spot) spectrum shows strong anisotropy of the emissivity
(limb-brightening effect) into account.

The source is characterized by $n_{\rm{} mean}$, the mean number of all
coexisting flares at a given time. The spots are assumed to appear randomly
across the disk surface, between the inner radius $R_{\rm{} in}$ (equal
to the marginally stable orbit around a black hole) and the adopted outer
radius, $R_{\rm{} out}$. The probability distribution of their radial and
azimuthal position, ($R_{\rm{}i},\phi_{\rm{}i}$), is determined by the general
law 

\begin{equation}
  p(R_{\rm{}i},\phi_{\rm{}i})\,{\equiv}\,
  p(R_{\rm{}i})\,p(\phi_{\rm{}i}) =  {(\gamma_{\rm rad} + 2)
  R_{\rm{}i}^{\gamma_{\rm rad} + 1} \over R_{\rm{}out}^{\gamma_{\rm rad}
  + 2}-R_{\rm{}in}^{\gamma_{\rm rad} +2}}\,{\times}\, {1 \over 2 \pi}.
  \label{eq:pr}
\end{equation}

This distribution reduces to the uniform distribution over the disk surface,
if $\gamma_{\rm rad} = 0$, while positive (negative) values of $\gamma_{\rm
rad}$ lead to an enhanced number of flares in the outer (inner) disk parts.

The spot radius, $R_{\rm X}$, is assumed to be the same for all spots. The
flare luminosity and, consequently, the local spot radiation flux generally
scales with the flare position, $R_{\rm{} i}$, as 

\begin{equation}
  F_{\rm{}inc} = F_0  \left({R_{\rm{}i}} \over
  {R_{\rm{}in}}\right)^{-\beta_{\rm{}rad}}.
  \label{eq:radial1}
\end{equation}

The dissipation within the Keplerian disk scales with $R^{-3}$, so the
typical value of $\beta_{\rm{}rad}$ is expected to be of that order.

The duration of the flare may also depend on the flare location, so we
describe it as

\begin{equation}
  t_{\rm life} = t_{\rm life_0} \left({R_{\rm{}i}} \over
  {R_{\rm{}in}}\right)^{\delta_{\rm{}rad}}.
  \label{eq:radial2}
\end{equation}

Here, $\delta_{\rm rad}=0$ means that the flare lifetime is independent
of the flare location, while for $\delta_{\rm rad}=1.5$, the flare lifetime
scales with the local Keplerian timescale.

The observational appearance of the source also depends on the duration of the
observation (i.e., integration time) and on the inclination angle of an
observer with respect to the disk symmetry axis. The value of the black hole
mass, $M$, is another relevant parameter since all distances and
timescales scale with $M$.

The flare distribution is fully described by the model parameters: $M$,
$n_{\rm{}mean}$, $R_{\rm{}in}$, $R_{\rm{}out}$,  $R_{\rm{}X}$, $F_0$,
$\beta_{\rm{}rad}$, $\gamma_{\rm{}rad}$, $\delta_{\rm{}rad}$, and
$t_{\rm{}life_0}$. The spot radius $R_{\rm{}X}$ can be replaced, if
convenient, by the mean X-ray luminosity of the source, $L_{\rm X}$.

\subsection{Modeling sequences of spectra}
\label{sect:sequences}

The observed properties of the spot distribution are strongly affected
by special and general relativistic effects. These effects must be taken into
account with the maximum possible accuracy since they account for a significant
fraction of the resulting variability. Even a stationary spot on a Keplerian
orbit gives a complex variability curve if the relativistic effects are
included. Light bending causes a magnification of the emission from a
relatively small disk area (located roughly behind the black hole). Flares
passing through this area strongly contribute to the rapid fluctuations
of the signal.

We apply the code {\sc KY} (see Dovciak et al. 2004 and Dovciak 2004 for a
description) here to compute the relativistic corrections between the local
spot spectra and the actually 'observed' ones. The code incorporates all
aspects of the light propagation from between the disk surface and a distant
observer. It includes the effects of light bending, as well as gravitational
and Doppler shifting of the photon energy. The ruling parameters for the
relativistic effects are the viewing angle $i$ of a distant observer, measured
with respect to the disk normal, and the dimensionless Kerr parameter $a$ of
the black hole. The anisotropic character of the local ($\psi$-dependent) spot
spectra (see Sect.~\ref{sect:integ}) is also relevant and included in our
modeling. 

The resulting disk spectrum seen at a given viewing angle $i$ is based on a
combination of various angle-dependent local spot spectra. However, the
integration over the disk surface for a large number of hot spots is
computationally time-consuming, so that we cannot produce detailed lightcurves
for all spots. However, to model the data, such lightcurves are
actually not needed because the spectra can only be measured in bins $T_{\rm
obs}$ of the typical order of a few thousands of seconds. Therefore, within
each time bin, we consider the appearance/disappearance of the hot spots and
their motion. Then we construct the emissivity `belts', which represent the
traces of the spots on the disk surface over their duration. This method is
identical to the one applied in Czerny et al. (2004). Finally, the spectrum
$S_{\rm{} k}(E)$ of the radiation received by a distant observer within a time
bin $k$, is computed. We calculate $N$ of such sequences for a given flare
distribution. Usually we adopt $N = 50$ since this corresponds well to the
best observed lightcurves covering a few hundreds of thousands of seconds.

Another modification, introduced in the current model in comparison to the
method of Czerny et al. (2004), is the survival of flares between consecutive
time bins. Czerny et al. (2004) treated each time bin as being separate from
the previous one, and all spot positions were newly generated. This is correct
if the time bins do not come from a single, uninterrupted observation or if
the effective life time of the flares is shorter than the length of a time
bin. In the current modeling, we follow the positions of the hot spots over
consecutive time bins. A randomly sampled position, according to an adopted
spot distribution, is assigned only to new born flares. This Monte-Carlo
sampling also considers an adopted birth rate, which is determined by the mean
number of flares, their life time, and by the spatial spot distribution. This
parameterization allows an appropriate description of the overall X-ray
properties of the source.

\subsection{Fractional variability}
  
The obtained radiation spectra sequences allow us to calculate the fractional
variability for a given flare distribution. We follow the definition used
by Ponti et al. (2004):

\begin{equation}
  {\hat F}_{\rm{} var}^2(E)=\sum_{k=1}^N {(S_{\rm{} k}(E)- \langle
  S(E) \rangle )^2 \over  (N-1) \langle S(E) \rangle^2},
\end{equation}

\noindent where $N$ is the number of time bins, $S_{\rm{} k}(E)$ is the
integrated radiation spectrum for a single time bin $k$, and $\langle S(E)
\rangle$ is the mean spectrum. We also determine the statistical errors for
this determination of ${\hat F}_{\rm{} var}$, as it is done for the observed
data (see Ponti et al. 2004). Of course the modeled lightcurve is not limited
by the S/N ratio, although it is limited by the adopted total duration of the
observation and by the width of the energy channels for the final $F_{\rm{}
var}$.

By statistical errors, we mean the error determination from a {\it single}
simulated lightcurve. In the simulations, we use a fine energy grid of 800
points in a 2 -- 10 keV energy range. The dimensionless ${\hat F}_{\rm{}
var}(E)$ is determined at each energy point. When we compare the model to
the data, we bin the results using broader energy bins. If $n$ original energy
points contribute to the new energy bin centered around $E_{\rm{} k}$, we
introduce

\begin{equation}
  F_{\rm{} var}(E_{\rm{} k}) = {\sum_{i=1}^{n} {\hat F}_{\rm{}
  var}(E_{\rm{} i}) \over n}, \label{eqn:fvar}
\end{equation}

\noindent and the error of this quantity is determined as

\begin{eqnarray}
  \delta F_{\rm{} var}(E_{\rm{} k}) & = & \left[ {\sum_{i=1}^{n}{\hat
    F}_{\rm{} var}^2(E_{\rm{} i})  - n F_{\rm{} var}^2(E_{\rm{} k}) \over n^2}
    + \right.\nonumber \\ &   & \left. {(F_{\rm{} var}^2(E_{\rm{} k}) +
    1)F_{\rm{} var}^2(E_{\rm{} k}) \over n (N-1)}\right]^{1/2}.
  \label{eqn:deltafvar}
\end{eqnarray}

\noindent The first term in this expression determines the error of the mean
value by averaging over the energy bins and assuming an error of ${\hat
F}_{\rm{} var}(E_{\rm{} i}) - F_{\rm{} var}(E_{\rm k})$ for a given ${\hat
F}_{\rm{} var}(E_{\rm{} i})$. The second term is the expected statistical error
for a normalized variance effectively calculated from $n(N-1)$ points. Both
components are necessary, since in the model the values in nearby energy bins
are not statistically independent.

\subsection{Point-to-point fractional variability}

We also calculate the point-to-point fractional variability (Edelson et
al. 2002; Vaughan \& Fabian 2004), which emphasizes the character of the
shortest timescale variation. This quantity is calculated by 

\begin{equation}
  {\hat F}_{\rm{} pp}^2(E)=\sum_{k=1}^{N-1} {4 \left[ S_{\rm{} k+1}(E)-S_{\rm{}
  k}(E) \right]^2 \over (N-1) \left[ S_{\rm{} k+1}(E)+S_{\rm{} k}(E)
  \right]^2},
\end{equation}

\noindent together with the rebinned version

\begin{equation}
  F_{\rm{} pp}(E_{\rm{} k}) = {\sum_{i=1}^{n} {\hat F}_{\rm{}
  pp}(E_{\rm{} i}) \over n}, \label{eqn:fpp}
\end{equation}

and the appropriate statistical errors $\delta F_{\rm{} pp}$. They are
computed from the same formula as $\delta F_{\rm{} var}$, replacing
${\hat F}_{\rm{} var}(E)$ by ${\hat F}_{\rm{} pp}(E)$ in
Eqn.~\ref{eqn:deltafvar} and $F_{\rm{} var}(E)$ by $F_{\rm{} pp}(E)$
respectively. Our definition of $F_{\rm{} pp}$ is even more local than the one
previously used because it adopts the local mean instead of a global mean at
each step.

\section{Results}

In our model setup, the vertical disk structure and the local spectra, as a
function of the emission angle $\psi$, are accurately computed for a distance
of 18 $R_{\rm{} g}$ from the black hole. These local spectra are then applied
to hot spots across the whole disk. On the one hand, this is an important
simplification in our modeling. On the other hand, however, it helps to
reduce the complexity of the problem. The model setup described in the
previous section focuses on investigating the effects on the variability
caused by General Relativity, by the radial spot distribution across the disk,
and by the anisotropic local spectrum emitted by the spots. The shape of the
local spectra for each spot are thus assumed to be independent of the spot
location, while their normalization is ruled by the radial emission profile
of the disk. Further radial dependencies, like the power of the primary
emission illuminating the disk or the hydrostatic equilibrium of the disk
atmosphere, are to be considered in future work.

The following presentation of our results starts with a comparison to the
previous approach of Czerny et al. (2004) in Sect.~\ref{sec:old-new}. Then,
in Sect.~\ref{sect:model_depen}, we point out the main effects on the
variability as a function of the model parameters, and we explain them
qualitatively. Finally, in Sect.~\ref{sec:mcg}, we apply the model to some
variability data of the Seyfert-1 galaxy MCG-6-30-15.

\subsection{The effect of the spot structure and of the anisotropic spot
  emission}
\label{sec:old-new}

\begin{figure}
  \epsfxsize=8.8cm
  \epsfbox{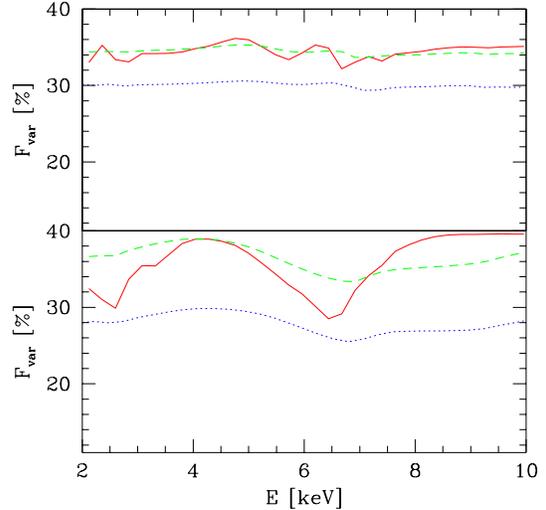}
  \caption{Fractional variability amplitude for a Schwarzschild black hole
    (upper panel) and a maximally rotating Kerr black hole (lower
    panel). Dotted lines: uniform, isotropically radiating spots generated
    independently in each time bin. Dashed lines: uniform, isotropically
    radiating spots surviving between consecutive time bins. Continuous lines:
    present model of non-uniform, non-isotropic spots surviving between
    consecutive time bins. Other model parameters: $M=10^8 M_{\odot}$, $L_{\rm
    X} = 10^{44}$ erg~s$^{-1}$, $i = 30^{\circ}$,  $R_{\rm{} out}=50 \;
    R_{\rm{} g}$, $\beta_{\rm{} rad} = 3$, $\gamma_{\rm{} rad} = 0$,
    $\delta_{\rm{} rad} = 0$, $t_{\rm{} life_0} = 10^5$ s, $T_{\rm{} obs} =
    10^5$ s, $N=300$; and $n_{\rm mean} = 30$ in the upper panel, while
    $n_{\rm mean} = 100$ in the lower panel. The inner radius is equal to the
    marginally stable orbit adapted to the given choice of the black hole
    spin. No primary contribution is assumed.}
  \label{fig:rms1}
\end{figure}

\begin{figure}
  \epsfxsize=8.8cm
  \epsfbox{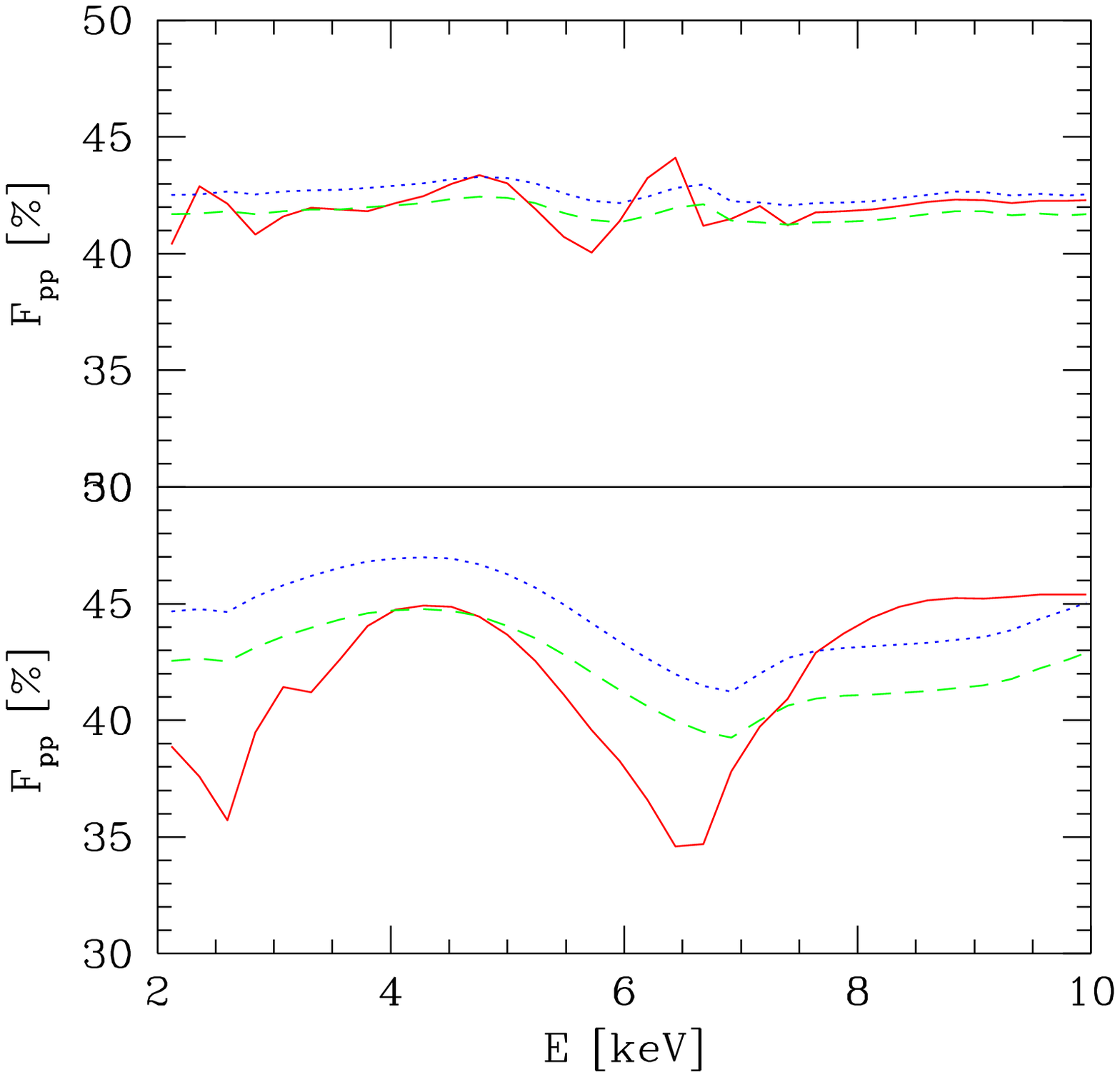}
  \caption{Point-to-point fractional variability amplitude for a Schwarzschild
    black hole (upper panel) and a maximally rotating Kerr black hole (lower
    panel). Dotted lines: uniform, isotropically radiating spots generated
    independently in each time bin. Dashed lines: uniform, isotropically
    radiating spots surviving between consecutive time bins. Continuous lines:
    present model of non-uniform, non-isotropic spots surviving between
    consecutive time bins. The other parameters of the models are as for
    Fig.~\ref{fig:rms1}.}
  \label{fig:rms2}
\end{figure}

Since we made two important modifications of the model in comparison to the
approach of Czerny et al. (2004), we show the significance of these changes in
Fig.~\ref{fig:rms1}, for two sets of parameters -- one corresponding to a
Schwarzschild black hole and one corresponding to a maximally rotating Kerr
black hole. The mean number $n_{\rm{} mean}$ of flares is adjusted so that
for both choices of the black hole spin, the variability level is roughly the
same (see Sect.~\ref{sect:model_depen}). We compare the results obtained
from (a) the code of Czerny et al. (2004), (b) the new code taking into
account the cloud survival between the time bins, but with the old spot
radiation spectrum, and (c) the new code, considering the spot structure and
the angle-dependent spot emission.

The differences between cases (a) and (b) are mainly in the normalization. The
survival of the flares between the consecutive time bins enhances the
variability since it decreases the randomness in the cloud position. The
discrepancy between the two approaches increases with a rising ratio of the
flare life time $t_{\rm{} life}$ to the bin size $T_{\rm{} obs}$. The
considered examples were calculated for $t_{\rm{} life}$ being independent
from the disk radius ($\delta_{\rm{} rad} = 0$). If the radial dependence is
strong (e.g. $\delta_{\rm{} rad} = 1.5$), the cloud survival also induces some
difference in the shape of $F_{\rm{} var}$.

The most profound difference is introduced by the change of the local
spectrum. Case (c) shows a much stronger energy dependence because the new
local reflection spectrum has more prominent spectral features, including the
iron $K\alpha$ line (see Sect.~\ref{sect:integ}). This is promising in the
perspective of reproducing the data. The model of Czerny et al. (2004)
predicted a far too shallow energy dependence, in comparison with observations
of MCG-6-30-15, whereas the new model improves this aspect. 

We also made a comparison of the predictions by the new version of the code
for the point-to-point variability. In this case, the cloud survival decreases
the variability amplitude since, for the adopted model parameters, about half
of the clouds survive from one time bin to another and do not contribute much
to the point-to-point variability. Some effect is also seen in the shape of
$F_{\rm{} pp}(E)$ (see Fig.~\ref{fig:rms2}).

Furthermore, we calculated $F_{\rm{} var}(E)$ and $F_{\rm{} pp}(E)$, taking
into account the contribution of the primary emission from the flare, with
a relative level, as expected for isotropic flare emission. In these cases, the
energy dependence of the variability is weak and comparable to the level
obtained by Czerny et al. (2004). Although in the new approach the energy
dependence in the spot emission variability is stronger, the spot spectrum
contributes less to the total (primary plus spot) spectrum due to the
less-ionized reflecting medium. The two effects roughly compensate for each
other and, for an isotropic primary, the variations in $F_{\rm{} var}(E)$ are
again around a mean variability level of about $2\%$.

\subsection{The dependence of the fractional variability on the model
  parameters} 
\label{sect:model_depen}

The mean number $n_{\rm{} mean}$ of flares mostly affects the overall
normalization of $F_{\rm{} var}$, as does the flare life-time distribution
$t_{\rm{} life}$: larger values of $t_{\rm{} life}$ require higher values of
$n_{\rm{} mean}$ to achieve the same overall variability level. The important
ruling parameter for the variability in this context is the frequency of
flare onsets and endings, which directly determines the fluctuations of the
lightcurve. If $n_{\rm{} mean}$ is high, more onsets and endings per
unit time happen and the averaging effect reduces the variability. The
lifetime of a single flare acts in the opposite direction. The longer an
individual flare lives, the less flares set on or end per unit time.

Suppression of $F_{\rm{} var}$ at $\sim 6.5$ keV appears if the radiation is
strongly concentrated toward the center. We see such a trend in models with
$\delta_{\rm{} rad}=0$, if the spot brightness radially decreases as the
standard disk dissipation rate ($\beta_{\rm{} rad}=3$) and the flares are
distributed uniformly across the disk surface ($\gamma_{\rm{} rad}=0$). We
show an example of this behavior in Fig.~\ref{fig:rms1} (continuous line).  If
$\beta_{\rm{} rad}$ is considerably larger (e.g. $\beta_{\rm{} rad} \sim 4.5
$), the same pattern is seen even for flare life times scaling with the local
Keplerian value ($\delta_{\rm{} rad}=1.5$). This behavior can be
qualitatively understood from the following considerations: the K$\alpha$ line
emitted from the innermost spots is relativistically broadened and
Doppler-shifted according to the local orbital velocity vector. Since the
life-time of a flare close to the marginally stable orbit is, in general, a
significant fraction of the orbital period, the line  emitted by one spot will
shift smoothly in energy around the line centroid. As the line is also
strongly relativistically broadened, its shape is rather flat around the
centroid, which hence exhibits less variability than do the much steeper
wings. In addition to that, there is a contribution to the iron K$\alpha$ line
from outer parts of the disk, and its persistent presence further supports the
drop in $F_{\rm{} var}$ around 6.5 keV. The features in the variability
spectrum are thus mainly determined by the relatively few innermost hot spots
for which the reprocessed spectrum is more strongly modified by the general
relativistic effects.

The parameters $\gamma_{\rm{} rad}$ and $\delta_{\rm{} rad}$ act predominantly
in the same way, since having more flares at larger distances is roughly
equivalent to flares lasting longer at large radii. However, the effect of the
two parameters is not strictly identical. The situation is similar with
$\beta_{\rm{} rad}$ and $\gamma_{\rm{} rad}$: a decrease in $\beta_{\rm{}
rad}$ has a similar effect as a decrease in $\gamma_{\rm{} rad}$. We should
add here that, to a certain extent, the timescale $t_{\rm{} life}$ also
influences the energy dependence of $F_{\rm{} var}$. If the timescale is too
short, the energy dependence in the variability is smeared, which is again due
to the large number of onsets and endings of the hot spots.

In summary, the most important property of a given parameter set is the sum
of the parameters, $ 2 + \gamma_{\rm{} rad} + \delta_{\rm{} rad} -
\beta_{\rm{} rad}$, which represents the index of an effective radial
distribution of energy generation. If this value is negative, the model is
likely to show a suppression of the variability around 6 keV. If the value is
positive, the model is likely to show complex behavior around 6 keV in
$F_{\rm{} var}$, as seen in Fig.~\ref{fig:random}, and an enhanced
variability at $\sim 6.5 $ keV in $F_{\rm{} pp}$, as seen in
Fig.~\ref{fig:random2}. However, to some extent the specific individual values
of all parameters are also important.

As mentioned above, the position of the features in $F_{\rm{} var}$ and their
shape strongly depend on the relativistic effects modifying the spectrum
of the inner hot spots. The observed features thus constrain both the
inclination angle $i$ of the disk and the Kerr black hole parameter, $a$. A
larger inclination tends to move the features toward higher energies, while a
faster black hole rotation produces slightly broader and much deeper
features. This is due to the effect that the marginally stable orbit of
a rotating black hole is smaller, and thus the hot spots can exist at a closer
distance to the event horizon. Therefore, if any features are present in
$F_{\rm{} var}$, they can be used effectively to constrain $a$ and $i$.

\subsection{Comparison to the special case of MCG-6-30-15}
\label{sec:mcg}

\subsubsection{Observational data and fixed model parameters}

Two long X-ray exposures were obtained with XMM-Newton for the Seyfert-1
galaxy MCG-6-30-15: one observation lasting 95 ks was performed in June 2000
(Wilms et al. 2001), and the second one lasting about 325 ks was done in July
2001 (Fabian et al. 2002). The time variability in the data from 2000 was
analyzed in detail by Ponti et al. (2004), while the time variability in the
data from 2001 was studied by Vaughan \& Fabian (2004).

We attempt to reproduce the shape of the fractional variability amplitude
obtained by Ponti et al. (2004; see their Fig.~2 and Fig.~\ref{fig:three} in
this paper). We also derive the point-to-point fractional variability
amplitude for the same data sequence. In our simulations, we adopt a fixed
value $T_{\rm{} obs} = 6146 $ s, as was done for the computations of $F_{\rm{}
var}$ during the data analysis by Ponti et al. (2004). The number of time
bins in the $F_{\rm{} var}$ calculations is $N = 16$, which is appropriate for
the data length. $F_{\rm{} pp}$ is calculated from the same data using 1000 s
bins, so in modeling $F_{\rm{} pp}$, we take $T_{\rm{} obs} = 1000 $ s, and the
number of points increases to $N=100$ to represent the data length
correctly. We adopt $10^7 M_{\odot}$ for the mass and $8 \times 10^{43}$ erg
s$^{-1}$ for the X-ray luminosity of the source. We fix the outer disk radius
somewhat arbitrarily at $R_{\rm{} out}=50 \; R_{\rm{} g}$. In most cases, we
assume $i=30^\circ$, a Kerr parameter of $a=0.95$, and $\gamma_{\rm{}
rad}=0$.

\subsubsection{Intrinsic randomness of the model}

\begin{figure}
  \epsfxsize=8.8cm
  \epsfbox{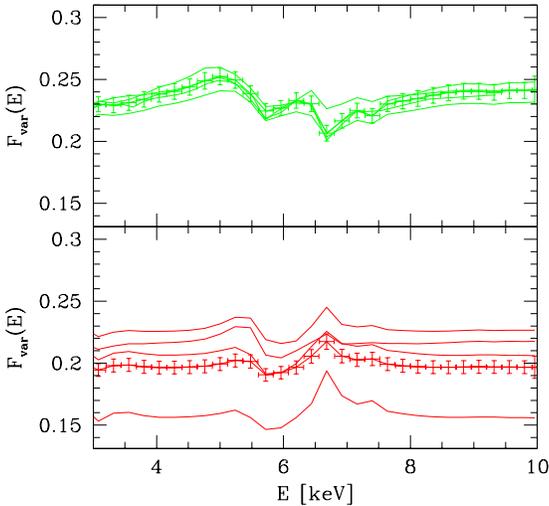}
  \caption{Fractional variability calculated from several lightcurves
    calculated for the same model, with the formal errors displayed for one of
    the curves. Upper panel: $n_{\rm{} mean}=125$, $\beta_{\rm{} rad}=3$;
    lower panel: $n_{\rm{} mean}=30$, $\beta_{\rm{} rad}=2$. Other model
    parameters: $M=10^7 M_{\odot}$, $a = 0.95$, $L_X = 8 \times 10^{43}$
    erg~s$^{-1}$, $i = 30^{\circ}$,  $R_{\rm{} out}=50 r_{g}$, $\gamma_{\rm{}
    rad} = 0$, $\delta_{\rm{} rad}= 1.5$, $t_{\rm{} life_0}= 2 \times 10^4$ s,
    $T_{\rm{} obs}=6146$ s, and $N = 50$. No primary contribution is assumed.}
  \label{fig:random}
\end{figure}

\begin{figure}
  \epsfxsize=8.8cm
  \epsfbox{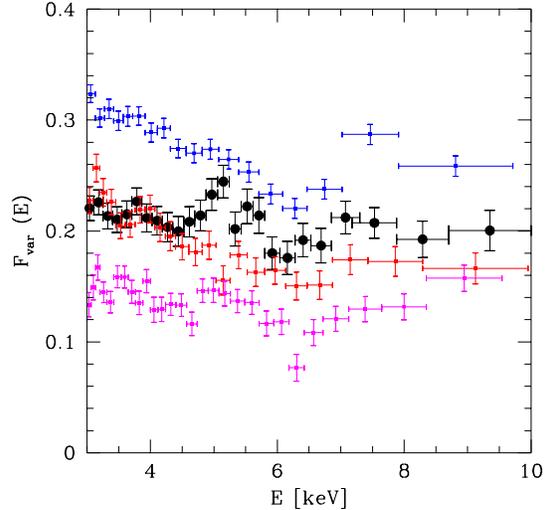}
  \caption{Fractional variability calculated independently for three separate
    parts of the 300 ks lightcurve of MCG-6-30-15 (XMM-Newton data, 2001,
    thin points). The set with filled circles represents the result obtained
    from the 100 ks lightcurve (XMM-Newton data, 2000). The length of a time
    bin equals 6146 s.}
  \label{fig:three}
\end{figure}

\begin{figure}
  \epsfxsize=8.8cm
  \epsfbox{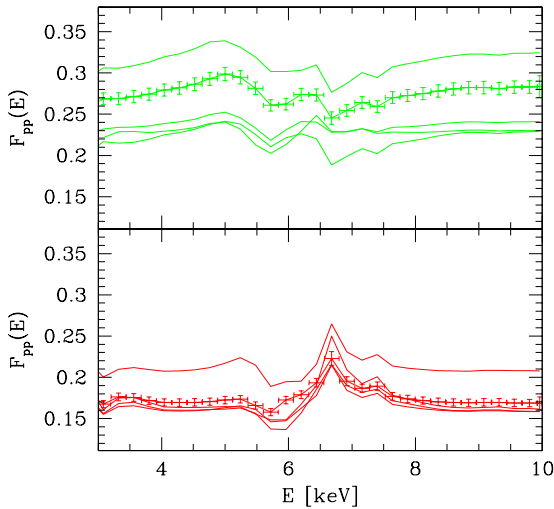}
  \caption{Point-to-point fractional variability calculated for the same
    models and displayed in the same manner as the fractional variability
    shown in Fig.~\ref{fig:random}.}
  \label{fig:random2}
\end{figure}

The most noticeable property of the model is the considerable randomness in
the results for some sets of model parameters and the relative stability for
others. The effect is shown in Fig.~\ref{fig:random}. We consider two
exemplary sets of parameters. In the first set (upper panel in
Fig.~\ref{fig:random}), the spot brightness decreased with the radius as
$\beta_{\rm{} rad} = 3$, in the second set with $\beta_{\rm{} rad} =
2$. Again, we choose the mean number of clouds in each model in such a way as
to have roughly the same level of variability. For a specific choice of the
other parameters (the same for the two models; see caption of
Fig.~\ref{fig:random}), we have to assume $n_{\rm{} mean}=125$ for the first
model and $n_{\rm{} mean}=30$ for the second one.

For both parameter sets, we obtained five 300-ks-long time sequences and
calculated $F_{\rm{} var}$ for each of the sequences independently. For
clarity, the expected statistical one sigma errors are marked for one of the
curves only. The curves in the upper panel are roughly similar, and they do
not differ more from each other than expected from simple error
analysis. However, the curves in the lower panel differ rather widely. This
demonstrates that, for such a set of parameters, a single lightcurve of a
standard length and with a standard error description cannot be used to
estimate whether the model well represents any specific data. The effect is
still somewhat stronger for our data covering only 100 ks instead of 300 ks,
as was adopted in these simulations.

The discrepancy is due to the different total mean numbers of flares required
by the two data sets. Since the flare duration distribution was the same in
the two models the total (time-integrated) number of flares in the lightcurve
is 726 for the first set of parameters and only 174 for the second set. The
standard determination of the statistical error (see Eqn.~A.2 of Ponti et
al. 2004 and Eqn.~\ref{eqn:deltafvar} in this paper) of course does not
include this effect. This result supports the conclusion of Vaughan et
al. (2003), based on Monte Carlo simulations of lightcurves with an assumed
power spectrum, that the overall normalization of the variance is rather
unreliable if based on short observations. However, our energy-dependent
calculations indicate that although the {\it overall level} of the $F_{\rm{}
  var}$ changes strongly for some sets of parameters, the {\it shape} of
$F_{\rm{} var}$ varies less from sequence to sequence. The energy pattern is
roughly preserved and can be used to estimate the source properties.

Since our conclusion is based on a specific flare/spot model, we compare the
trend to the sampling effects in real data: in Fig.~\ref{fig:three} we show
three $F_{\rm{} var}(E)$ plots obtained by the division of a single 300 ks long
lightcurve of MCG-6-30-15 (XMM-Newton observation performed in 2000, see
Fabian et al. 2002) into three equal parts.  $F_{\rm{} var}(E)$ was calculated
for each of the three parts independently. The overall change in the
variability level between the three subsequences is in good agreement with
the expected power red leak for the source. Table 1 of Vaughan et al. (2003)
for a source with a PSD slope of 1.0 below the break and 2.0 above the break,
for 20 data points,  predicts the scatter between -0.71 and 0.45 in logarithm
of the variance, so between  -0.36 to 0.23 in logarithm of dispersion. This
means that the average value of 0.2 is predicted to fluctuate between 0.09 and
0.33, as observed (see Fig.~\ref{fig:three}). However, most of the variability
is in the overall normalization of $F_{\rm{} var}(E)$; the shape actually
varies much less, although some changes in energy dependence are also
seen. The variability  is always suppressed just above 6 keV, but one of the
data sets does not show the complex structure in 4 - 6 keV band.

The structure of the $F_{\rm{} var}(E)$ spectrum of MCG-6-30-15 shown
in Fig.~\ref{fig:three} is quite typical for Seyfert galaxies. The suppression
around 6 keV is often found to be even more pronounced than for this
particular object (see, e.g., Markowitz, Edelson, \& Vaughan, 2003, for
some examples). Another important feature of the rms spectra is an
increase of the variability in the soft X-ray band. An interesting
interpretation of this feature is proposed by Chevallier et
al. (2006). According to this idea, the enhanced soft X-ray
variability is due to the response of the warm absorber to the
variations of the incident flux. Absorption is most efficient
around $\sim 1$ keV and the change of the ionization state of the warm
absorber in the response to the variable flux amplifies the overall
variability level in this spectral band. This leads to an increase in
the simulated fractional variability amplitude at about that energy.

\begin{figure}
  \epsfxsize=8.8cm
  \epsfbox{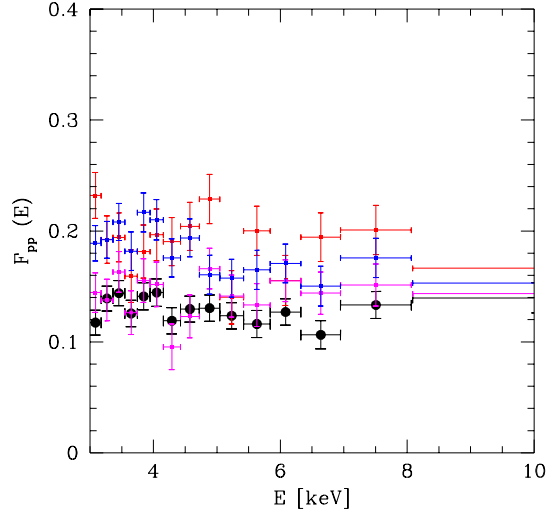}
  \caption{Point-to-point fractional variability calculated independently for
    three separate parts of the 100 ks lightcurve of MCG-6-30-15 (XMM-Newton
    data, 2000, thin points). The set with filled circles represents the mean
    result obtained directly from the whole lightcurve. The length of a time
    bin equals 1000 s.}
  \label{fig:three_pp}
\end{figure}

We also analyze the scatter in point-to-point fractional variability amplitude
predicted by the model. The result is shown in Fig.~\ref{fig:random2}. It is
interesting to note that the results are complementary to the results for the
simple fractional variability. While in Fig.~\ref{fig:random} there is much
more scatter in the upper panel, in Fig.~\ref{fig:random2} the trend is
reversed. This complementarity is connected with the presence of relatively
more (or less) power in a given model at longer timescales. Also, in this case,
the model predicts much less variability in the shape than in the overall
normalization. In the observed $F_{\rm{} pp}$, calculated separately for three
parts of the lightcurve, we notice considerable variations in the
normalization, although lower than for $F_{\rm{} var}$ (see
Fig.~\ref{fig:three_pp}). We see some variations of the shape as well, but the
basic pattern remains the same. There is a decrease of the variability at
$\sim 6$ keV, which is present in all three sequences, although in one of them
it is shifted from 6.6 keV to 6.0 keV. It is difficult to discuss the
significance of such variations since the errors are large, although the
number of bins was relatively high due to the choice of short time bin (1000
s).

This analysis shows that the observational errors are actually much larger
than indicated by a simple statistical analysis of a given single lightcurve,
as is expected from the power leaking (Vaughan et al. 2003). However, the
errors in the shape are not as large as the overall change of fractional
variability amplitude from one to another time sequence, due to the long
timescales present in the system. Of course, an increase of the duration of
the observation time would give less scatter. But then, such requirements may
be difficult to meet. The model from the lower panel of Fig.~\ref{fig:random}
gives results for $F_{\rm{} var}$ that are almost within the statistical
errors of a single lightcurve, if the number of time bins is increased to
$N=300$. This corresponds to an observation of more than 20 days.

\subsubsection{Best representations of $F_{\rm{} pp}(E)$}

\begin{figure}
  \epsfxsize=8.8cm
  \epsfbox{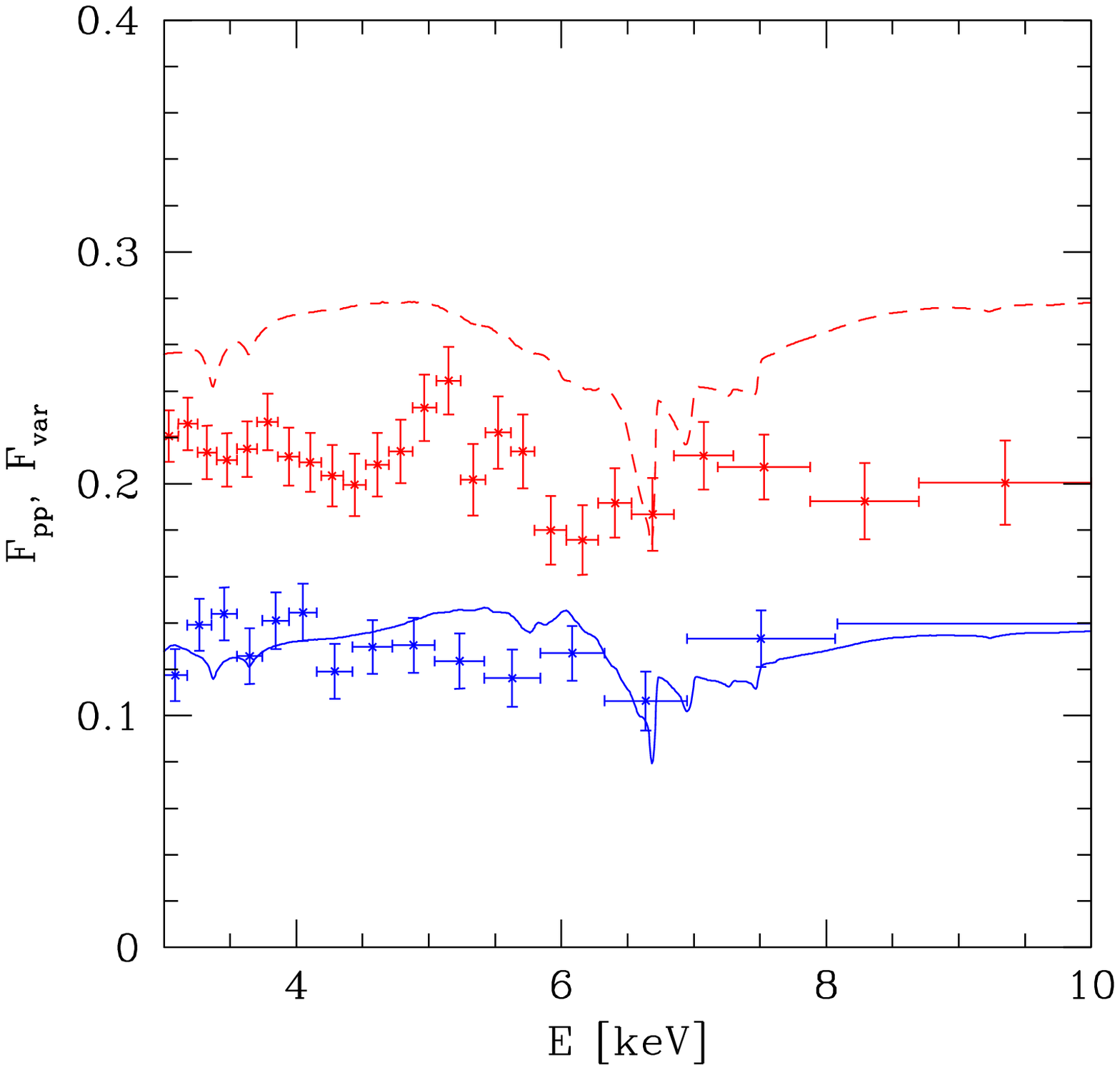}
  \caption{The best model of $F_{\rm{} pp}$ (continuous line) with the
    corresponding data (lower set of points). The upper data set represents
    $F_{\rm{} var}$ and the modeling result for the same parameter set:
    $\beta_{\rm{} rad} = 4$, $\gamma_{\rm{} rad} = 0$, $\delta_{\rm{} rad}=
    1.5$, $t_{\rm{} life_0}= 2 \times 10^4$ s, and $n_{\rm mean}=750$. The
    other parameters are at their standard values (see caption of
    Fig.~\ref{fig:rms1}). The model is shown unbinned for better clarity.}
  \label{fig:best_short}
\end{figure}

In Fig.~\ref{fig:best_short}, we show the point-to-point fractional variability
amplitude of MCG-6-30-15 and the best representation of the short timescale
variability from our model. The model well represents the suppression of the
variability near the core of the iron line, in the 3-10 keV energy band. The
position of the minimum in the theoretical plot is basically determined by the
inclination angle, and apparently, the value of 30$^{\circ}$ adopted in most
simulations is appropriate. Inclinations smaller than $\sim 10^{\circ}$ or
larger than $\sim 40^{\circ}$ are excluded.

The width of the dip is mostly determined  by the Kerr metric parameter
$a$. Fast rotation of the black hole results in a broad feature extending far
to the red side of the line core, while a slowly rotating black hole gives a
narrow feature. The value $a=0.95$, relatively high, but still significantly
lower than the maximally rotating black hole, best represents the data. A
slower rotation than $a \sim 0.8$ is not consistent with the data. The depth
of the feature mostly depends on the ratio of the spot spectrum to the primary
emission going towards an observer. Our best representation was obtained
assuming no primary contribution, but the contribution of the primary at some
level is allowed, as analyzed in detail in Sect.~\ref{sect:prim}. The overall
level of the variability depends mostly on the mean number of flares, and this
can always be adjusted if the energy-dependent shape of the $F_{\rm var}$ is
appropriate. However, the same model parameters do not represent the $F_{\rm{}
var}$ well.

\subsubsection{Best representations of $F_{\rm{} var}(E)$}

  \begin{figure}
  \epsfxsize=8.8cm
  \epsfbox{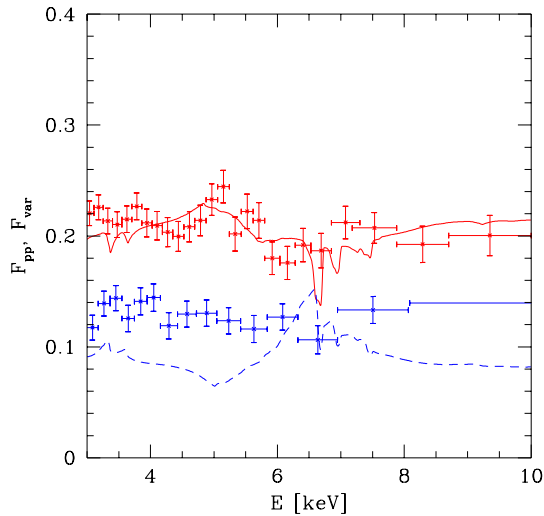}
  \caption{The best model of $F_{\rm{} var}$ (continuous line) with the
    corresponding data (upper set of points). The lower data set represents
    $F_{\rm{} pp}$ and the modeling result for the same parameters:
    $\beta_{\rm{} rad} = 3$, $\gamma_{\rm{} rad} = 0$, $\delta_{\rm{} rad} =
    1.5$, $t_{\rm{} life_0} = 2 \times 10^5$ s, and $n_{\rm mean} = 100$. The
    other parameters are at their standard values (see caption of
    Fig.~\ref{fig:rms1}). The model is shown unbinned for better clarity.}
  \label{fig:best_long}
\end{figure}

In Fig.~\ref{fig:best_long}, we show the best representation of the standard
$F_{\rm var}(E)$ for the  source. It roughly reproduces the observed `wiggles'
of the $F_{\rm{} var}$ obtained from the whole lightcurve. However, this time
the same parameters do not reproduce $F_{\rm{} pp}$ correctly. Since there is
a considerable randomness in the modeling process, we repeated the computations
several times for a different random seed of the Monte-Carlo routines. But in
all such realizations, the $F_{\rm{} pp}$ shows an excess at 6.4 keV instead of
a dip. A discrepancy with the data therefore remains.

\begin{figure}
  \epsfxsize=8.8cm
  \epsfbox{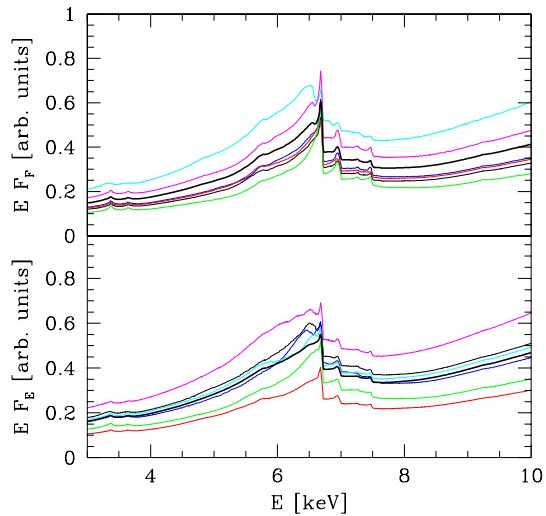}
  \caption{The spectra at various moments computed from the models shown in
    Fig.~\ref{fig:best_short} (lower panel) and Fig.~\ref{fig:best_long}
    (upper panel), with the mean spectra over-plotted by a thick line.}
  \label{fig:two_spec}
\end{figure}

\begin{figure}
  \epsfxsize=8.8cm
  \epsfbox{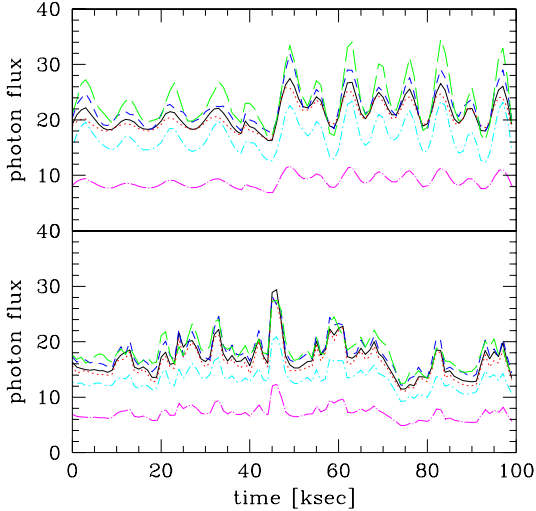}
  \caption{The lightcurves of the models shown in Fig.~\ref{fig:best_short}
    (lower panel) and Fig.~\ref{fig:best_long} (upper panel) for narrow energy
    bands around 4.4 keV, 5.2 keV, 6 keV, 6.4 keV, 6.8 keV, and 8.8 keV, and
    with the resolution used for the $F_{\rm{} pp}$ computations.}
  \label{fig:two_curve}
\end{figure}

To decide which set of parameters better represents the source properties, we
plot the spectra and the lightcurves from the two models (see
Fig.~\ref{fig:two_spec}). The spectra of both models are quite comparable and
do not help to differentiate between the candidate models. The lightcurves, on
the other hand, have a rather different character visually (see
Fig.~\ref{fig:two_curve}). The upper curve shows less short time-scale
flickering and seems to be quasi-periodic, while the lower curve looks more
similar to the source X-ray lightcurve (see Fig.~1 in Ponti et al. 2004). 

Therefore, we consider the set of parameters used for
Fig.~\ref{fig:best_short} as a more satisfactory description of the
variability in MCG-6-30-15. 

\subsubsection{Contribution of the primary emission}
\label{sect:prim}

To directly compute observable characteristics, we need to combine the primary
continuum with the reflected component. The resulting signal depends on the
mutual proportion of the two components, and also on the energy range and time
resolution of the observation. This means that the degree of anisotropy of the
primary flare should be taken into account. Our modeling shows that, if the
primary power law emission strongly dominates over the reprocessed component,
and if the slope of the primary emission does not change with time, there is
no energy-dependence in the fractional variability amplitude. Moderate
contribution of the primary results in flattening of the energy dependence
without changing the overall variability amplitude. We note, though, that this
is only true for the case of a well-defined correlation between the primary
and the reprocessed component. This correlation implies that the flare
geometry shown in Fig.~\ref{fig:rings} is the same for all flares distributed
over the disk. In reality, these things are likely to be more complicated. If
the primary emission and the reprocessing are not perfectly correlated, the
overall level of the variability goes down  when the primary component becomes
more significant.

We calculate examples of the model shown in Fig.~\ref{fig:best_short} with
different levels of primary contribution and assuming that both components are
well-correlated. We also calculate the equivalent width of the iron line in
the resulting spectra. These results are shown in Table~\ref{tab:ew}. The
model without any primary contribution (anisotropy parameter ${\mathcal R} =
\infty$) best represents the decrease of the variability at $\sim 6.6 $ keV,
but the EW of the iron line complex in the predicted mean spectra is very
large, as can be seen in Fig.~\ref{fig:two_spec}. The reflection amplitude
${\mathcal R} \sim 4$ is still roughly acceptable from the point of view of
the variability. The actual measurements of the line EW in MCG-6-30-15 gave
values ranging from 800 eV to 250 eV, depending on the assumption about the
underlying continuum (Wilms et al. 2001), and Reynolds et al. (2004) obtained
the value of $500^{+300}_{-200}$ eV, so the line intensity is not strongly
overestimated. The model with ${\mathcal R} = 2$ seems to have too low of an
energy dependence in the fractional variability amplitude, and for ${\mathcal
R} = 1$ it underestimates the line intensity as well.

\begin{table}
  \caption{Equivalent widths (EW) of the iron line complex and the ratio
    $F_{\rm{} pp}$(6 keV)/$F_{\rm{} pp}$(6.6 keV) as a function of the amount
    of reflection. The observed $F_{\rm{} pp}$(6 keV)/$F_{\rm{} pp}$(6.6 keV)
    ratio is 1.19, and the model results were binned into the same energy
    channels as the data.
  \label{tab:ew}}
  \begin{center}
    \begin{tabular}{lrrr}
      \hline
      {$\mathcal R$} & $EW^a$[eV] & $EW^b$[eV]& $F_{\rm{} pp}$(6 keV)/
      $F_{\rm{} pp}$ (6.6 keV)\\
      \hline
      $\infty$   & 1670  & 1350 & 1.17 \\
      4          &  660  &  510 & 1.14 \\
      2          &  420  &  320 & 1.09 \\
      1          &  240  &  180 & 1.06 \\
      \hline
    \end{tabular}
  \end{center}
  $^a$ measured in the 3 - 8 keV range, $^b$ measured in the 4 - 9 keV range,
  with the continuum determined outside the iron line complex.
\end{table}

The amount of reflection can be estimated from the strength of the Compton
hump seen in hard X-ray data. The analysis of Beppo-SAX data by Ballantyne et
al. (2003) showed that the reflection factor determined from {\sc pexrav} was
$R=2.7^{+1.4}_{-0.9}$ for solar abundances. Assuming higher iron abundance
than solar improved the fit and only gave a lower limit to the reflection
factor ($R > 2.6$) so the spectrum can  actually be reflection-dominated, as
advocated for NLS1  (including MCG-6-30-15) by Fabian et al. (2002). A purely
reflection-dominated spectrum ($R > 3$) for MCG-6-30-15 was also considered by
Taylor, Uttley, \& McHardy (2003).  Therefore, from the point of view of the
primary contribution, the favored model is satisfactory.

\section{Discussion}
\label{sect:discussion}

The X-ray spectra of a number of AGN contain a reflection feature
which is best modeled as a relativistically smeared iron line (for a
review, see, e.g., Reynolds \& Nowak 2003).  If the source of the
irradiation consists of multiple localized flares, it is natural to expect the
variability of a single spot luminosity (e.g. Abramowicz et al. 1991;
Bao 1992; Karas et al. 1992), as well as of the line shape (Ruszkowski
2000; Nayakshin \& Kazanas 2001; Turner et al. 2002; Iwasawa, Miniutti, \&
Fabian 2004).

In the present paper, we studied in detail the time-dependent spectra from a
multi-flare model, paying particular attention to the energy-dependent
fractional variability amplitude as a diagnostic tool. We used both the
standard fractional variability amplitude, $F_{\rm{} var}$, and the
point-to-point fractional variability amplitude, $F_{\rm{} pp}$. We showed
that the model explains a decrease in the variability level at $\sim 6.5 $ keV
derived from the rms-spectra of Seyfert galaxies, such as MCG-6-30-15 (Inoue
\& Matsumoto, 2001; Vaughan \& Fabian, 2004; Ponti et al., 2004). Similar
behavior was observed in Mrk~766 (Pounds et al., 2003), 3C~120, Akn~564,
IC~4329a, 3C~390.3, Fairall~9, NGC~3516, NGC~3783, NGC~4151, NGC~5548, or
NGC~7469 (Markowitz, Edelson, \& Vaughan, 2003).

The overall variability level, predicted by the multi-flare model and
expressed in terms of $F_{\rm{} var}$ and $F_{\rm{} pp}$, depends strongly on
the adopted number of flare-spots. By adjusting this number, the model is
capable of reproducing the overall normalization of observed rms-spectra as we
demonstrated for the case of MCG-6-30-15. However, the form of the energy
dependence of $F_{\rm{} var}$ and $F_{\rm{} pp}$ is more difficult to adjust
and provides a strong diagnostic tool either for proof or for  falsification
of our model.

The prediction of the multi-flare model for the shape of $F_{\rm{} var}$ and
$F_{\rm{} pp}$ depends predominantly on the radial profile of the energy
dissipation characterizing a given set of parameters  (see
Sect.~\ref{sect:model_depen}). If the energy dissipation in the  form
of flares weighted within the emission area decreases with the radius,
the model shows a suppression of the variability at 6.6 keV. 

The present data predict both $F_{\rm{} var}$ and $F_{\rm{} pp}$ with
significant error, mostly due to the presence of the timescales much longer
than a typical lightcurve duration (e.g. Vaughan et al. 2003). The errors in
the shape of the energy dependence seem to be lower, but still rather difficult
to determine observationally. Therefore, we did not provide formal fits to the
observational data  at the present stage. Instead, we favored one of the
studied sets of  parameters  at the basis of semi-quantitative analysis.

\subsection{Best model properties for MCG-6-30-15}

The parameters of the model which we considered satisfactory (see
Fig.~\ref{fig:best_short}) are interesting. The distribution of the
timescales was Keplerian, and the requested normalization leads to a
timescale range from 1910 s in the innermost part to $8.7 \times 10^4$
s in the outermost part. The spot size obtained was $R_{\rm{} X} =
1.96 \; R_{\rm{} g}$, which is not too large in the context of the 
underlying model assumptions. It corresponds to a height of the flare source
above the disk equal to $1.13 \; R_{\rm{} g}$. The mean number of coexisting
flares is quite large (750), but the variability is nevertheless observed to
be that large since most of the luminosity comes only from the few innermost
flares. This is due to the rather steep increase of the flare luminosity
toward the disk center ($\beta_{\rm{} rad}=4$). Interestingly, the obtained
mean number of flares is not unlike the numbers of clouds obtained in the
obscurational variability model of Abrassart \& Czerny (2000).  

The steep flare emissivity mentioned above is indeed higher than might
be expected just from the shape of the gravitational
potential. However, modeling of the mean shape of the broad iron line
frequently required an enhanced emissivity profile (e.g. Wilms 
et al. 2001). It was also predicted by a theoretical corona model of
Kawanaka et al. (2005).

Although the mean number of the coexisting flares is large, we see
occasional large `single' flares in the theoretical lightcurve. Such
resolved single flares are already seen occasionally in the present
data (e.g.  Ponti et al. 2004 for MCG-6-30-15). Turner et al. (2005)
identified a single orbiting flare in Mrk~766 on the basis of the
spectral analysis, and Porquet et al. (2004) also considered a
localized flare as a plausible explanation of the ESO 113-G010
spectrum.

\subsection{Model approximations}

There are still certain inconsistencies present in our model,
mainly in the computation of the local spectra. The calculations were
made for a Schwarzschild black hole of $10^8$ solar masses. However,
in computing the relativistic effects, we also use them for other
values of $M$. For MCG-6-30-15, a smaller mass is more appropriate. In
our computations of the relativistic effects, we assumed $10^7
M_{\odot}$ for the mass of the central black hole, and McHardy et
al. (2005) suggested an even smaller value $\sim (3 - 6) \times 10^6
M_{\odot}$). However, we do not expect any strong direct dependence of the
local spectrum on the black hole mass.

The local computations were performed at a single radius ($18 R_{\rm{}
g}$), but we checked that the local spectra calculated at $7 R_{\rm{}
g}$ do not differ strongly (Goosmann et al. 2006) when computed for
the same flare-to-disk-flux ratio below the flare. Therefore, for
models calculated with this ratio preserved (i.e., models with
$\beta_{\rm{} rad}=3$) we do not think we introduce much of an
error. However, for models with different $\beta_{\rm{} rad}$ the
local spectrum may be different, so a range of local models should actually be
calculated to incorporate this effect.

The assumptions about the primary radiation in our model are rather
simple. The distance between the flare source and the hot spot is assumed to
be negligibly small so that the relativistic effects acting on the incident
radiation can be neglected. In Goosmann et al. (2006), we show that this is a
good approximation for Schwarzschild black holes. For Kerr black holes it
works fine if the flares do not get closer to the black hole than $\sim 3
R_{\rm{} g}$. The primary source is assumed to turn on and off instantaneously
and, consequently, so are the hot spots. We thereby assume that the time for
the actual reprocessing of the radiation in the disk medium is negligible. The
local lightcurve of the spot emission hence has a simple box shape. We keep
this assumption, although it is probably too simple, since not much is known
yet about the actual production of the primary radiation and its time
evolution during a magnetic reconnection event. In later developments of the
model, more detailed assumptions about the nature of the flare source should
be included. Also the ratio of the primary flux to the disk flux is somehow
arbitrarily set to 144 at the spot center. In future work, this flux ratio
should be treated as a free parameter, but of course this requires a large
grid of calculations for the local spectrum.

\subsection{Power spectrum}

The power spectrum derived by McHardy et al. (2005) for MCG-6-30-15
shows a bending, characterized by the break frequency $3.2 \times
10^{-5}$ Hz, and the asymptotic slopes of -0.8 and -1.74 if the broken
power law description is used. It is not clear that our model is able
to reproduce its properties. We attempted to calculate the power
spectrum from our theoretical lightcurves, but the result was
consistent with white noise. This may be partially due to the fact
that: (i) the lightcurves contained only a limited  number of points
(100), and (ii) the studied dynamical range of the model ($R_{\rm{} out} = 50
\; R_{\rm{} g}$) is not yet fully satisfactory.

On the other hand, our model in its present version may not contain enough
power in the long frequency range.  Generally, more complex models involving
correlated avalanches of flares are used to model the power spectra in AGN and
galactic sources (e.g. Poutanen \& Fabian 1999; Zycki 2003; Zycki \&
Niedzwiecki 2005). A successful model accounting for the power spectrum has to
explain why there is a lot of power in the low frequency part of the power
spectrum, while most of the energy dissipation takes place in the innermost
part where the expected timescales are short.

Two essentially similar models were proposed which can explain the
presence of the long timescales successfully: 'self-organized
criticality' (Mineshige et al. 1994), and 'traveling
perturbations', (Lyubarskii 1997). Both conclude that the disk
accretion rate is locally perturbed, and this perturbation propagates
inward, so the accretion rate at any disk radius contains a history of
long timescale perturbations created at larger radii. In particular, the
model of Lyubarskii (1997) looks natural within the frame of the
magneto-rotational instability mechanism of viscosity, and it is
particularly appropriate for a power spectrum with a single break like
that of MCG-6-30-15.

Our flare model, at present, does not include this kind of
perturbation memory. However, in principle, this kind of non-local
perturbation of the disk accretion rate can be incorporated into the
model. If it is  combined with some assumptions of scaling between the
local momentary accretion rate and a flare luminosity, it will
probably provide the missing power at long timescales.

\subsection{Other models of X-ray variability}

We have found that our multi-flare model with a time-independent
shape of the energy spectrum of the flare emission provides an
interesting interpretation of the drop in the variability around 6.5
keV. However,  there are other models under discussion.

The energy-dependent form of the variability was also modeled as a
variable  primary emission with constant reflection (e.g. Taylor,
Uttley, \& McHardy, 2003; Shih, Iwasawa, \& Fabian, 2002), or as
a result of the pivoting of the primary power law emission
(e.g. Markowitz, Edelson, \& Vaughan, 2003).

However, in this first class of models it is difficult to explain why the
reflection component should be constant since either it forms close to the
black hole (and the source of the primary emission) and should respond to
changes of the primary, or it forms far away, then it is difficult to explain
why it is relativistically broadened. An attractive way out of this paradox
was proposed by Martocchia, Karas, \& Matt (2000); Miniutti \& Fabian (2004);
and Miniutti et al. (2005), who describe a model where the change in the
primary intensity is predominantly due to the minor changes in the position of
the source of the primary radiation at the disk rotation axis, which gave
effects strongly enhanced due to the  general relativity. Pivoting is also
quite natural, if the flare evolution (cooling, expansion) is taken into
account. However, in this case the drop in the variability level can be at any
position, and not just around the iron line complex.

There are also entirely different models of the X-ray variability of AGN, for
example the presence of the density inhomogeneities in the disk  atmosphere
resulting from the photon bubble instability (Ballantyne et al. 2005),
developments of the temporary spiral structures in  the accretion disk
(Fukumura \& Tsuruta 2005), variations in the properties of the warm absorber
(e.g. Sako et al. 2002; Done \& Gierlinski, 2006, private communication), or
variations in the the inner disk torque (Garofalo \& Reynolds 2005). Only more
detailed studies of model predictions and comparison with the data in more
than just one aspect may bring a reliable answer to the true nature of the
X-ray variability.

\begin{acknowledgements}

Part of this work was supported by the grants PBZ-KBN-054/P03/2001 and 4 T12E
047 27 of the Polish State Committee for Scientific Research, by the
Laboratoire Europ\'een Associ\' e Astrophysique Pologne-France, by the CNRS
GDR PCHE, by the Hans-B\"ockler-Stiftung in Germany (RWG), and by the Center
of Theoretical Astrophysics in Czechia. The authors gratefully acknowledge
support from the Czech Science Foundation  GACR 202/06/0041 (VK) and
205/05/P525 (MD). The Astronomical Institute in Prague has  been operated
under the project AV0Z10030501. GP thanks the European commission under the
Marie Curie Early Stage Research Training program for support.

\end{acknowledgements}

\end{document}